\documentstyle[twoside,fleqn,espcrc2]{article}
\newfont{\liste}{pzdr scaled 1100}
\newfont{\grfett}{cmmib10 scaled 1100}
\newcommand{\lsim}{\mbox{\raisebox{-0.3ex}{%
\footnotesize $\:\stackrel{<}{\sim}\:$}} }

\newcommand{\AmS}{{\protect\the\textfont2
  A\kern-.1667em\lower.5ex\hbox{M}\kern-.125emS}}
\newcommand{\MZ}{M_Z}
\newcommand{\az}{\alpha(\MZ)}
\newcommand{\aiz}{\alpha^{-1}(\MZ)}
\newcommand{\bea}{\begin{eqnarray}}
\newcommand{\eea}{\end{eqnarray}}
\newcommand{\be}{\begin{eqnarray*}}
\newcommand{\ba}{\begin{eqnarray*}}
\newcommand{\ee}{\end{eqnarray*}}
\newcommand{\ea}{\end{eqnarray*}}
\newcommand{\bary}{\begin{array}}
\newcommand{\eary}{\end{array}}
\newcommand{\bit}{\begin{itemize}}
\newcommand{\eit}{\end{itemize}}
\newcommand{\dalf}{\Delta\alpha}
\newcommand{\das}{\Delta\alpha(s)}

\newcommand{\dahs}{\Delta\alpha^{(5)}_{hadrons}(s)}
\newcommand{\dahz}{\Delta\alpha^{(5)}_{hadrons}(\MZ^2)}
\newcommand{\nn}{\nonumber}
\newcommand{\crn}{\nn \\}
\newcommand{\noi}{\noindent}
\newcommand{\ra}{\rightarrow}
\newcommand{\al}{\alpha}

\newcommand{\veps}{\varepsilon}
\newcommand{\epm}{e^+e^-}
\newcommand{\mz}{M_Z^2}
\newcommand{\bi}[1]{\bibitem{#1}}
\newcommand{\ci}[1]{\cite{#1}}
\newcommand{\dal}{\Delta \alpha}
\newcommand{\dalh}{\Delta \alpha^{(5)}_{\rm had}}
\newcommand{\npb}[1]{{Nucl.\ Phys.\ }{B{#1}} }
\newcommand{\np}[1]{{Nucl.\ Phys.\ }{B{#1}} }
\newcommand{\plb}[1]{{Phys.\ Lett.\ }{B{#1}} }       % Vol 171 ...
\newcommand{\pl}[1]{{Phys.\ Lett.\ }{{#1}B} }        % Vol 1 to 170
\newcommand{\prd}[1]{{Phys.\ Rev.\ }{D{#1}} }
\newcommand{\PR}[1]{{Phys.\ Rep.\ }{C{#1}} }
\newcommand{\prl}[1]{{Phys.\ Rev.\ Lett.\ }{{#1}} }
\newcommand{\pr}[1]{{Phys.\ Rev.\ }{{#1}} }
\newcommand{\zp}[1]{{Z.\ Phys.\ }{C{#1}} }

\title{Hadronic Vacuum Polarization Contribution to \\  $g-2$
of the Leptons and $\alpha(M_Z)$}

\author{F. Jegerlehner\address{DESY-IfH Zeuthen,
                               Platanenallee 6, D-15738 Zeuthen, Germany}% 
        \thanks{Report on work in collaboration with
        S. Eidelman~[1].}}

\begin{document}
\onecolumn{
\renewcommand{\thefootnote}{\fnsymbol{footnote}}
\setlength{\baselineskip}{0.52cm}
\thispagestyle{empty}
\begin{flushleft}
DESY 96--121 \\
June 1996\\
%hep-ph/9606484\\
\end{flushleft}

\setcounter{page}{0}

\mbox{}
\vspace*{\fill}
\begin{center}
{\Large\bf Hadronic Vacuum Polarization Contribution} \\  
\vspace{3mm}
{\Large\bf  to $g-2$ of the Leptons and $\alpha(M_Z)$} \\

\vspace{5em}
\large
F. Jegerlehner\footnote[4]{\noindent Report on work in collaboration with
        S. Eidelman. To appear in the Proceedings of the Workshop
        ``QCD and QED in Higher Orders'', Rheinsberg, Germany, 1996,
        Nucl. Phys. B (Proc.Suppl.) to appear} 
\\
\vspace{5em}
\normalsize
{\it   DESY--Zeuthen}\\
{\it   Platanenallee 6, D--15738 Zeuthen, Germany}\\
\end{center}
\vspace*{\fill}}
\newpage
\begin{abstract}
We review and compare recent calculations of hadronic vacuum
polarization effects. In particular, we consider the anomalous magnetic
moments $g-2$ of the leptons and $\alpha(M_Z)$ , the effective fine
structure constant at the $Z$--resonance.
\end{abstract}

\maketitle

\section{VACUUM POLARIZATION AND CHARGE SCREENING}

\begin{figure}[h]
\begin{picture}(10,50)
\includegraphics{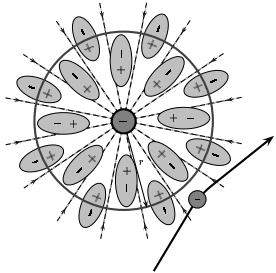}
\end{picture}
\caption{Vacuum polarization by virtual pair creation.}
\label{fig:fig1}
\end{figure}

Typically, charged particles in a collision of impact energy $E$
interact electromagnetically with an effective charge which is the
charge contained in the sphere of radius $r\simeq 1/E$ around the
particles.  As illustrated in Fig.~\ref{fig:fig1} for one of the
particles, the effective charge, due to vacuum polarization by virtual
pair--creation, is larger than the classical charge which is seen in a
large sphere ($r \rightarrow \infty$). This {\em charge screening} is a
particular kind of charge renormalization.\\

\vspace*{-7mm}

\begin{figure}[h]
\begin{picture}(10,81)
\includegraphics{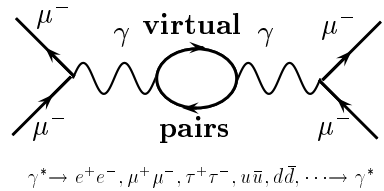}
\end{picture}
\caption{Feynman diagram describing the vacuum polarization in muon scattering.}
\label{fig:fig2}
\end{figure}

Commonly, Fig.~\ref{fig:fig1} is represented by a Feynman diagram like
the one in Fig.~\ref{fig:fig2} contributing to muon scattering.
Not surprising, the effective fine structure ``constant'' $\alpha(E)$
appears in many places in physics whenever the typical energy of a
process is not in the classical regime. The major contribution to
charge screening comes from light charged particle--antiparticle pairs
of mass $m \lsim E/2$. While the lepton contributions can be easily
calculated in QED perturbation theory the contribution of the strongly
interacting quarks is not so easy to obtain. This is the issue of our
discussion.

\vspace*{-7mm}

\vspace*{-7mm}

\subsection{Formal definition:}

\vspace*{3mm}

The effective QED coupling constant at scale $\sqrt{s}$ may be written as
\bea
\alpha(s) =  \frac{\alpha}{1 - \das}
\eea
with
\bea
\das =- 4\pi\alpha
        {\rm Re}\left[ \Pi'_{\gamma}(s) - \Pi'_{\gamma}(0)
                \right]
\eea
where $\Pi'_{\gamma}(s)$ is the photon vacuum polarization function
\bea
&& i\int d^4x\,e^{iq\cdot x}\langle 0| {\rm T} j_{em}^\mu(x) j_{em}^\nu(0)
                        |0 \rangle
\crn &=& -(q^2 g^{\mu\nu} - q^\mu q^\nu)\Pi'_{\gamma}(q^2)
\eea
and $j_{em}^\mu(x)$ is the electromagnetic current.
\subsection{Contributions:}

\vspace*{4mm}

{\large $\bullet$}~~{\large {Leptons}} \\

Their contribution is calculable in perturbation theory. The free
lepton loops are affected by small electromagnetic corrections only. In
leading order one obtains:
\footnotesize
\be \bary{l}
\dalf_{{\rm leptons}}(s)= \cr \bary{lcl}
& = & \sum\limits_{\ell=e,\mu,\tau}
      \frac{\alpha}{3\pi}
      \left[ - \frac{8}{3} + \beta_\ell^2
             - \frac{1}{2}\beta_\ell(3 - \beta_\ell^2)
               \ln\left( \frac{1-\beta_\ell}{1+\beta_\ell}
                  \right)
      \right]   \cr
& = &
      \sum\limits_{\ell=e,\mu,\tau}
      \frac{\alpha}{3\pi}
      \left[ \ln\left( s/m_\ell^2
                \right)
           - \frac{5}{3}
           + O\left( m_\ell^2/s
              \right)
      \right] ~~~~~~{\large (4)}   \\
& = & 0.03142 {\rm \ for \ } s=M_Z^2 \eary \eary
\label{eq:leptons}
\ee
\setcounter{equation}{4}
\normalsize
where $\beta_\ell = \sqrt{1 - 4m_\ell^2/s}$.\\

{\large $\bullet$}~~{\large {Quarks}}\\

The contribution of the five light quarks ($u,d,s,c,b$) is not
calculable in perturbation theory. The free quark loops are strongly
modified by strong interactions at low energy. The way out is the
following: unitarity and the analyticity of $\Pi'_{\gamma}(s)$ allow
us to write {(Cabbibo, Gatto 1961\ci{CabbiboGatto61})}
\bea
\dahs
 = - \frac{\alpha s}{3\pi}\;{\cal P}\!\!\! 
   \int_{4m_\pi^2}^\infty ds' \frac{R(s')}{s'(s'-s)}
\label{eq:DI}
\eea
where
\be
R(s) \equiv \frac{\sigma(e^+e^- \rightarrow \gamma^* \rightarrow hadrons)}{
\frac{4\pi\alpha^2(s)}{3s}}
= 12\pi{\rm Im}\Pi'_{\gamma}(s)
\ee
has been measured in $e^+e^-$--annihilation experiments up to energies
above which we may calculate it in perturbative QCD. In other words\\

{$\bullet$}~~$R(s)$ in known for low and moderately high $s$ from the
measurement of the total hadronic cross--section $e^+e^- \rightarrow
{\rm hadrons}$. This is of particular importance in the low energy
region and for the resonance regions where non-perturbative physics
comes into play.\\

{$\bullet$}~~$R(s)$ may be reliably calculated for large $s$ in {\em
perturbative QCD} by virtue of the asymptotic freedom of QCD. The
leading contribution is given by the sum over the squares of the
charges $Q_q$ of all quarks $q$
\be
R(s) \simeq N_c\:\sum\limits_{q} Q_q^2\:(1+O(\alpha_s/\pi))
\ee
with $N_c=3$ the color factor.\\

We thus may evaluate the hadronic part of the vacuum polarization by
utilizing $e^+e^-$--data (non--perturbative) for $\sqrt{s} \lsim
E_{{\rm cut}}\sim 40 {\rm GeV}$ plus the perturbative tail. Of course
the data exhibit experimental uncertainties which will allow us to
estimate this contribution with limited accuracy only.

Before we continue to discuss the evaluation of the dispersion
integral~(\ref{eq:DI}), let us remind the reader that the dispersion
integral representation derives from basic properties valid for
any quantum field theory:\\

\begin{figure}[h]
\begin{picture}(10,10)
\includegraphics{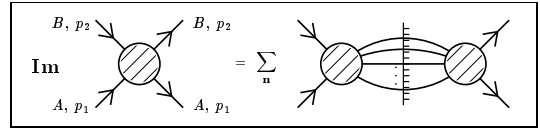}
\end{picture}
\caption{Optical theorem for scattering.}
\label{fig:optthscat}
\end{figure}

{\Large \liste\symbol{'157}}~~{\em Unitarity} implies the {\em
optical theorem:} The imaginary part of the forward scattering
amplitude of an elastic process $ A+B \ra A+B $ is proportional to the
sum over all possible final states $A+B \ra $ ``anything'' (see
Fig.~\ref{fig:optthscat})

\vspace*{-4mm}

\be \bary{c}
{\rm \ Im \ }{\rm T}_{{\rm forward}}\:\left(A+B \ra A+B \right) =\\
\sqrt{\lambda \left( s,m_1^2, m_2^2 \right)} \:
\sigma_{tot}\left(A+B \ra {\rm \ anything} \right) \eary
\ee

\vspace*{-4mm}

\vspace*{-6mm}

\begin{figure}[h]
\begin{picture}(10,10)
\includegraphics{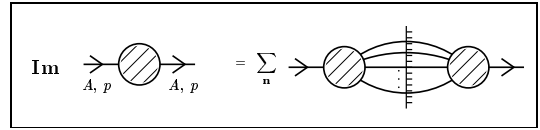}
\end{picture}
\caption{Optical theorem for propagation.}
\label{fig:optthprop}
\end{figure}

The corresponding relation for the photon propagator reads (see
Fig.~\ref{fig:optthprop})

\vspace*{-4mm}

\be
{\rm Im} \Pi'_{\gamma}(s)=\frac{1}{12 \pi} R(s)
\ee

\vspace*{-4mm}

\vspace*{-6mm}

{\Large \liste\symbol{'157}}
{\em Causality} implies {\em analyticity} which may be expressed in
form of a so--called (subtracted) dispersion relation
\be
\Pi'_{\gamma}(q^2)-\Pi'_{\gamma}(0)=
\frac{q^2}{\pi}\int\limits_{0}^{\infty} ds
\frac{{\rm Im} \Pi'_{\gamma}(s)}{s\:(s-q^2-i\veps)}\;\;.
\ee
The latter, together with the optical theorem, directly implies the
validity of (\ref{eq:DI}). Note that its validity is based on general
principles and holds beyond perturbation theory. It is the basis of
all non--perturbative evaluations of hadronic vacuum polarization effects
in terms of experimental data.
\subsection{Description of the $e^+ e^-$--data.}

\vspace*{3mm}

In a recent reanalysis~\ci{EJ95} of the hadronic vacuum polarization
we have collected all published $e^+ e^-$--data plus some unpublished
data (see also \ci{Landolt}). The data sets obtained for different
energy regions have been displayed in a number of figures in
Ref.~\ci{EJ95}, and we will not repeat them here. The following {\bf
integration procedure} has been used for the evaluation of the
dispersion integral:
\begin{itemize}
\item [{\Large \liste\symbol{'266}}]
Take data as they are and apply trapezoidal rule (connecting data
points by straight lines) for integration.
\item [{\Large \liste\symbol{'267}}]
To combine results from different experiments: i) integrate data
for individual experiments and combine the results, ii) combine data from
different experiments before integration and integrate combined
``integrand''. Check consistency of the two possible procedures to
estimate the reliability of the results.

\item [{\Large \liste\symbol{'270}}]
Error analysis: 1) statistical errors are added in quadrature, 2)
systematic errors are added linearly for different experiments, 3)
combined results are obtained by taking weighted averages. 4) all
errors are added in quadrature for ``independent'' data sets. We
assume this to be allowed in particular for different energy regions
and/or different accelerators.
\item [{\Large \liste\symbol{'271}}]
Resonances have been parametrized by Breit--Wigner shapes with
parameters taken from the Particle Data Tables~\ci{PDG}.
\end{itemize}
Our recent update {(Eidelman, Jegerlehner 1995)}
\be
\dahz
 = 0.0280 \pm 0.0007
\ee
together with the leptonic term~(\ref{eq:leptons}) yields an

\newpage

\begin{table}
\caption[]{Comparison of estimates of $\dalh (\mz)$}
\begin{tabular}{ll}
\hline
$\dalh (\mz)$ & Author (Year) [Ref.] \\
\hline
0.0285~(7) & Jegerlehner      (86)    \ci{Jegerlehner86} \\
0.0283(12) & Lynn et al.      (87)    \ci{LynnPensoVerz} \\
0.0287~(9) & Burkhardt et al. (89)    \ci{Burkhardt89} \\
0.0282~(9) & Jegerlehner      (91)    \ci{TASI} \\
0.0273~(4) & {Martin, Zeppenfeld} (94)  \ci{MZ94} \\
0.0280~(7) & {Eidelman, Jegerlehner} (95)  \ci{EJ95}\\
0.0280~(7) & {Burkhardt, Pietrzyk} (95)  \ci{BP95}\\
0.0275~(5) & Swartz           (95)    \ci{Swartz96} \\
0.0289~(4) & {Adel,Yndur\'ain}  (95)    \ci{AY95}\\
\hline
\end{tabular}
\label{tab:dalh}
\end{table}

\vspace*{-6mm}

\noi
effective fine structure constant at the $Z$--resonance:$$\aiz=128.89 \pm
0.09\;\;.$$

\begin{table*}
\caption[]{Contributions to $\dalh \times 10^4$ and relative(rel) and
absolute(abs) errors in percent.}
\begin{tabular}{ccr|crcc}
\hline
 final state & range (GeV) & contribution~~~~~~ &
& contribution &rel \%&abs \%\\
\hline
  $ \rho   $     &(0.28, 0.81)      & 26.08 (0.29) (0.62) &&&&\\
  $ \omega $     &(0.42, 0.81)      &  2.93 (0.04) (0.08) &&&&\\
  $ \phi   $     &(1.00, 1.04)      &  5.08 (0.07) (0.12) &
          Resonances:           &   46.61 (1.08) &   2.3  &   0.4  \\
  $ J/\psi $     &                  & 11.34 (0.55) (0.61) &&&&\\
  $\Upsilon$     &                  &  1.18 (0.05) (0.06) &&&&\\
  hadrons        &(0.81, 1.40)      & 13.83 (0.15) (0.79) &Background: &&&\\
  hadrons        &(1.40, 3.10)      & 27.62 (0.32) (4.01) &
{\small     $E<M_{J/\psi}$ }    &   41.45 (4.11) &   9.9  &   1.5  \\
  hadrons        &(3.10, 3.60)      &  5.82 (0.30) (1.12) &
{\footnotesize $M_{J/\psi}<E<3.6$ GeV} &    5.82 (1.16) &  19.9  &   0.4  \\
  hadrons        &(3.60, 9.46)      & 50.60 (0.24) (3.33) &
{\footnotesize $3.6$ GeV $<E<M_{\Upsilon}$}&   50.60 (3.34) &  6.6  & 1.2 \\
  hadrons        &(9.46, 40.0)      & 93.07 (0.86) (3.39) &
{\footnotesize $M_{\Upsilon}<E<$   40 GeV}&   93.07 (3.50) &   3.8  & 1.2  \\
    QCD   &(40.0, $\infty$ )   & 42.82 (0.00) (0.10) &
  {\footnotesize 40 GeV $ < E$ QCD}         &   42.82 (0.10) & 0.2  & 0.0  \\
\hline
   total       &                    &280.37(1.18) (6.43) &
     total                &  280.4 (6.54) &\multicolumn{2}{c}{ 2.3 } \\
\hline
\end{tabular}
\end{table*}
\subsection{Results for $\az$:}

\vspace*{3mm}

\begin{figure}[h]
\begin{picture}(10,10)
\includegraphics{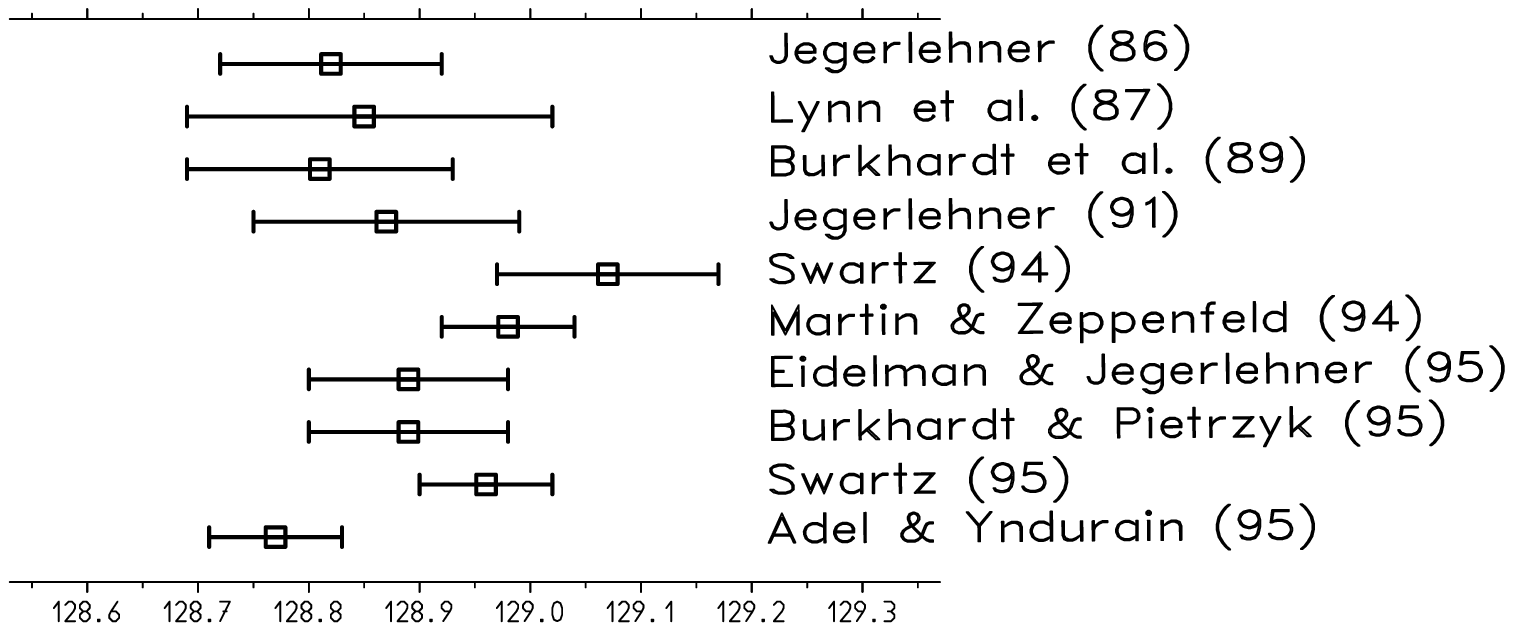}
\put(0,-34){\framebox(210,120)}
\end{picture}

\vspace*{2mm}

\caption{Different estimates of $\az ^{-1}$.}
\label{fig:alpstat}
\end{figure}

In Table~\ref{tab:dalh} and Fig.~\ref{fig:alpstat} we
show a comparison of results obtained by different authors. All values
have been rescaled to $M_Z=91.1887$ GeV. The small top quark and the
$W$ contributions are not included in $\az$.
For a comparison of the earlier results~\ci{Berends76} we refer to
\ci{Jegerlehner86}.  The differences are mainly due to a different
treatment of the systematic errors and/or different model assumptions.
Detail of our analysis are given in Table~2. Note that a reduction
of the error would require a precision measurement of the $\sigma_{\rm
tot}(e^+e^- \rightarrow {\rm hadrons})$ from 1 GeV to about 10 GeV.

\vspace*{2cm}

\vspace*{-7mm}

A few remarks concerning the results presented in Tab.~\ref{tab:dalh}
are in order here. Updates seemed to be justified from time to time
when new experimental data became available. While there have been very
little truly new experimental results, some groups have published
updated results which were available before as preliminary data
only. Examples are the ND results~\ci{Dolinsky91} up to 1.4 GeV, the
DM2 results~\ci{Schioppa86} between 1.35 and 2.3 GeV and the Crystal
Ball results~\ci{Edwards90} between 5.0 and 7.4 GeV. New results from
VEPP-4~\ci{Blinov91}, from 7.2 to 9.5 GeV, have been included as
well. In addition in
\ci{EJ95} we have made an effort to collect as much as possible all
the available data.

Additional motivations for performing the update \ci{EJ95} were the
following: the recent issues of the Review of Particle
Properties~\ci{PDG} had some improvements on the resonance parameters
and there was progress in calculating $R(s)$ at high energies in
perturbative QCD. For the muon anomaly the study of the low energy end
by means of chiral perturbation theory allowed us to reduce potential
model-dependences of earlier approaches.

The update \ci{Burkhardt89} attempted to estimate an ``official''
number which could be used by LEP experiments in their analyses. As
compared to \ci{Jegerlehner86}, a more conservative guess for the
uncertainty in the continuum above the $\rho$ up to the $J/\Psi$ was
made, and therefore, the uncertainty increased from 0.0007 to
0.0009. Later, the Crystal Ball (CB) collaboration carefully
reanalyzed their old $\epm$- annihilation data and obtained $R(s)$
values substantially lower than the Mark I data and in agreement with
other experiments (LENA). The results were in much better agreement
with perturbative QCD.  The change of the data was mainly due to an
up-to-date treatment of the QED radiative corrections and
$\tau$--subtraction. In Ref.~\ci{TASI} I included the updated results
from CB and discarded the Mark I data, which systematically lie 28\%
higher, on average. Note that, because of the much larger systematic
uncertainties of the Mark I data, the results are affected in a minor
way if we exclude (50.60(0.24)(3.33)) or include (50.79(0.20)(3.20))
the Mark I data. The inclusion of the CB data lead to a reduction of
previous results by -0.0005. While in
Refs.~\ci{Jegerlehner86,Burkhardt89,TASI} the $\rho$ region was
parametrized by an improved Gounaris--Sakurai formula, in our last
update
\ci{EJ95} we apply trapezoidal rule integration and include the final 
data from VEPP-2M OLYA, ND and CMD up to 1.4 GeV. This extended
analysis confirmed our previous results. From 1.35 to 2.3 GeV new data
from ORSAY DCI DM2 lead to a reduction of the error in this
region. The error now agrees with the earlier estimate in
\ci{Jegerlehner86}.

While the analysis \ci{BP95} is similar in spirit to ours and
reproduced our result, the estimate \ci{MZ94} is biased by believing
in perturbative QCD down to 3 GeV; some experimental data used in the
$J/\Psi$ and $\Upsilon$ resonance regions are rescaled using
perturbative QCD. In contrast to this estimate which yields a low
value, the estimate \ci{AY95}, which also relies on perturbative QCD,
finds a large value; both estimates result in much lower errors than
what can be justified by the data alone. In the analysis of
\ci{Swartz96} data are fitted before integration. This requires
a guess of the functional form of the integrand which is then
fitted to the data. In addition one has to assume some kind of
correlation matrix between the data points. One consequence of the
method applied is that the inclusion of one additional unpublished
data point~\ci{Osterheld86} led to a shift of the result by 1 $\sigma$.

\begin{figure}[h]
\begin{picture}(10,10)
\includegraphics{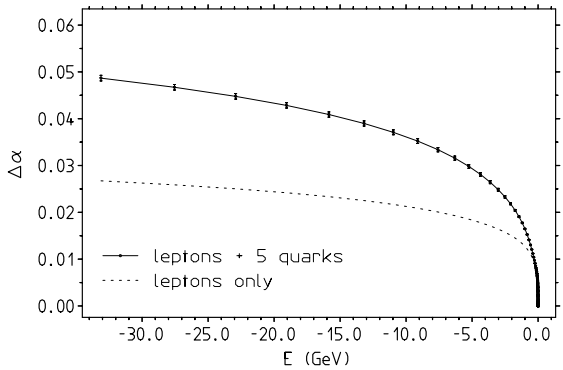}
\end{picture}

\vspace*{12mm}

\caption{$\Delta \alpha(-Q^2)$ in the spacelike region.}
\label{fig:alphaslow}
\end{figure}

In Fig.~\ref{fig:alphaslow} we show the
result for $\alpha(-Q^2)$ as a function of $E=|Q|$ for low $t=-Q^2$ in
the spacelike region, which is relevant for the $t$--channel
contribution to Bhabha scattering.  Note the\\

\vspace*{16.5mm}

\vspace*{1.2cm}

\begin{figure}[th]
\begin{picture}(10,10)
\includegraphics{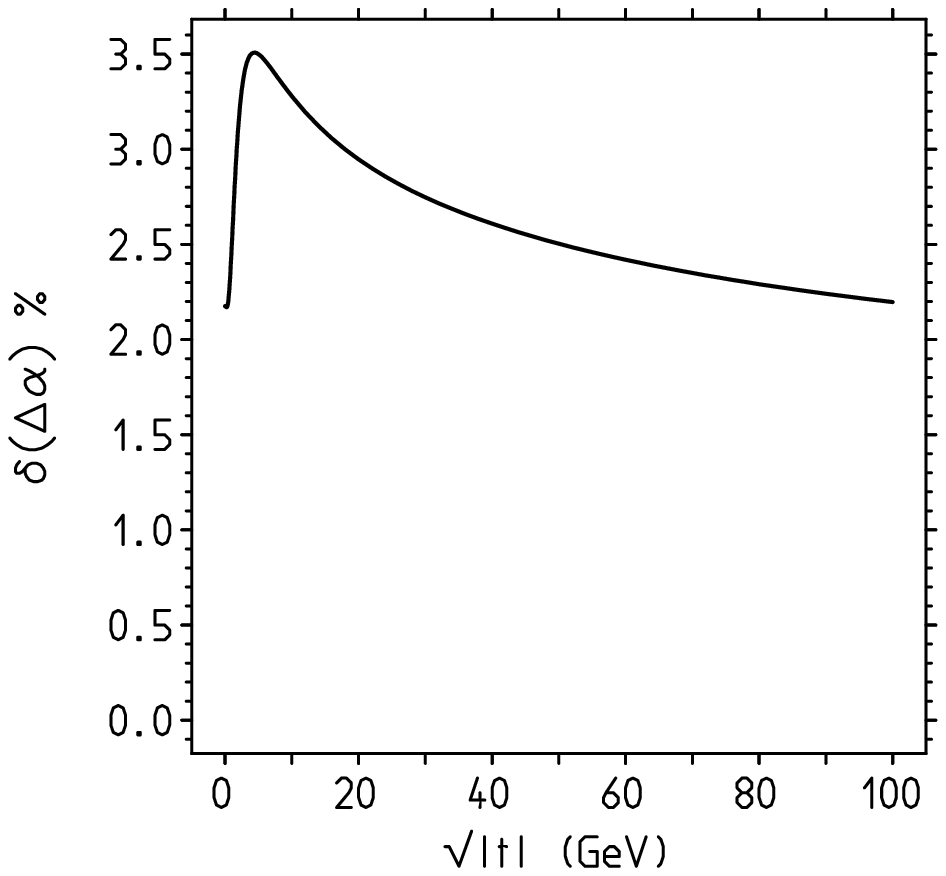}
\includegraphics{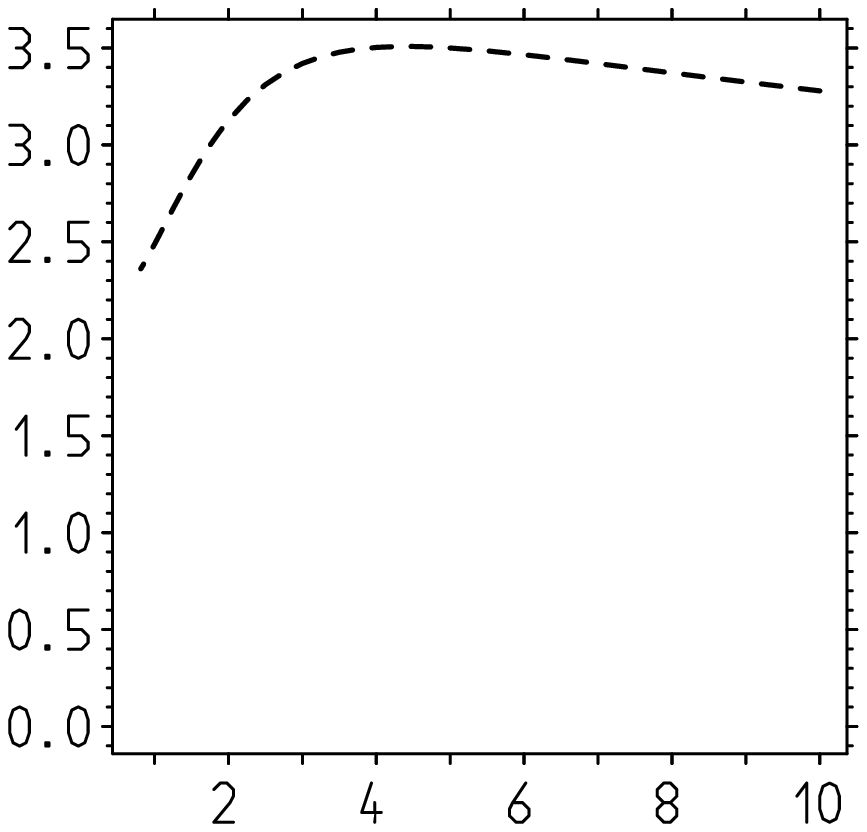}
\end{picture}

\vspace*{15mm}

\caption{Uncertainty of $\Delta \alpha(t)\;$.}
\label{fig:bhaerr}
\end{figure}

\vspace*{-5mm}

\noi
dramatic increase of the effective charge at low spacelike
momenta. This shows that the classical limit is difficult to reach in
a scattering experiment. A clean measurement of the running of
$\alpha(-Q^2)$ is possible at LEP by an appropriate analysis of the
small angle Bhabha scattering data. The values of $Q^2$ should be low
enough such that the $t$--channel contribution is clearly dominant.
In Fig.~\ref{fig:bhaerr} we show the uncertainty in \% of our
evaluation of the running charge in the low energy region.
\subsection{Parametrizations.}

\vspace*{3mm}

We have performed fits of $\dalh(s)$. The
parametrizations\footnote{Fortran routines {\bf
\normalsize hadr5.f} and {\bf \normalsize alphaQED.f} for calculating
$\dalh (s)$ and $\alpha (s)$, respectively, are available from WWW
http://www.ifh.de/$\tilde{~~}$fjeger/ } and the
best fit parameters we found are the following: \\
\noi
For $-(2 {\rm GeV})^2 < s < (0.25 {\rm GeV})^2$ the best fit is
\be
\dalh (s) &=& c_1\:\left\{\ln|1-c_2s|+c_2\frac{c_3s}{c_3-s}\right\}\\&&
-l_1 \frac{c_3s}{c_3-s}+c_4\:(s/s_0)^2
\ee
with $s_0 =   -(2 {\rm GeV})^2$,
     $l_1=  9.3055 \times 10^{-3}$ GeV$^{-2}$,\\ \noi
\begin{tabular}{ll}
     $c_1=  2.2694240 \times 10^{-3}$,&
    $\!\! c_2=   8.073429 {\rm GeV}^{-2}$,\\
     $c_3=  0.1636393 {\rm GeV}^{2}$,&
    $\!\! c_4= -3.35455 \times 10^{-5}$.\\
\end{tabular}
\noi
In the range  $-(20 {\rm GeV})^2 < s < -(2 {\rm GeV})^2$ we find
\be
\dalh (s) &=& c_1\:\left\{ \ln |1-c_2s|+c_2 \frac{c_3s}{c_3-s}\right\} \\&&
-l_1 \frac{c_3s}{c_3-s}+c_4 \: (s/s_0)
\ee
with $s_0 =  -(20 {\rm GeV})^2$,
     $l_1=  9.3055 \times 10^{-3}$ GeV$^{-2}$,\\ \noi
\begin{tabular}{ll}
     $c_1=  2.8668314 \times 10^{-3}$,&
    $\!\! c_2=  0.3514608 {\rm GeV}^{-2}$,\\
     $c_3=  0.5496359 {\rm GeV}^{2}$,&
    $\!\! c_4=  1.989233 \times 10^{-4}$.\\
\end{tabular}
At higher energies, in the ranges $E_{\rm min}< E < E_{\rm max}$, we have
\be
\dalh (s) &=& c_1+c_2 \: ( \ln (s/s_0) \\&&
+ c_3\:(s_0/s-1)+c_4\:((s_0/s)^2-1))
\ee
where $s= E |E|$ and $s=E_0 |E_0|$ with

\vspace*{5mm}
\noi
\begin{tabular}{crrrrr}
\hline
\hline
$E_{\rm min}$      &  250     &  80      &     40   \\
$E_{\rm max}$      & 1000     & 250      &     80   \\
$E_0$              & 1000     & 91.1888  &     80   \\
\hline
$c_1\times 10^{2}$ & 4.2092260& 2.8039809& 2.7266588\\
$c_2\times 10^{3}$ & 2.9233438& 2.9373798& 2.9285045\\
$c_3\times 10^{3}$ &-0.3296691&-2.8432352&-4.7720564\\
$c_4\times 10^{4}$ & 0.0034324&-5.2537734& 7.7295507\\
\hline
\hline
$E_{\rm min}$      &  -200    &  -1000   &\\
$E_{\rm max}$      &   -20    &   -200   &\\
$E_0$              &  -100    &   -1000  &\\
\hline
$c_1\times 10^{2}$ & 2.8526291& 4.2069394&\\
$c_2\times 10^{3}$ & 2.9520725& 2.9253566&\\
$c_3\times 10^{3}$ &-2.7906310&-0.6778245&\\
$c_4\times 10^{4}$ & 0.6417453& 0.0932141&\\
\hline
\end{tabular}

\vspace*{5mm}

At the higher energies, $E \geq 40$ GeV, we may fit the hadronic
contribution using the following {\sl effective parameters} in a
$O(\alpha_s)$ perturbative QCD formula: effective quark masses:
$m_{u,d,s,c,b}=0.067,0.089,0.231,1.299,4.500$ GeV together with an
effective $\alpha_s=0.102$. Thus
\be
\dalh (s) = - \frac{\alpha}{9\pi}\:
(1+\alpha_s^{\rm eff}/\pi)\:~~~~~~~~~~~~~~~~~~~~~\\
~~~~~~~(h(y_d)+h(y_s)+h(y_b)+4 (h(y_u)+h(y_c)))
\ee
where {\small
\be
h(y)&=&5/3+y-(1+y/2)\: g(y)\\
g(y)&=&\left\{ \bary{ccc}  2 \sqrt{y-1} \arctan(1/\sqrt{y-1}) & {\rm
      for} & y>1 \\ \sqrt{1-y}\ln(|\frac{1+\sqrt{1-y}}{1-\sqrt{1-y}}|) &
      {\rm for} & y<1 \eary \right.
\ee }
with $y_i=4 m_i^2/s$.

\section{VACUUM POLARIZATION CONTRIBUTION TO $g-2$}

\begin{figure}[h]
\begin{picture}(10,60)
\includegraphics{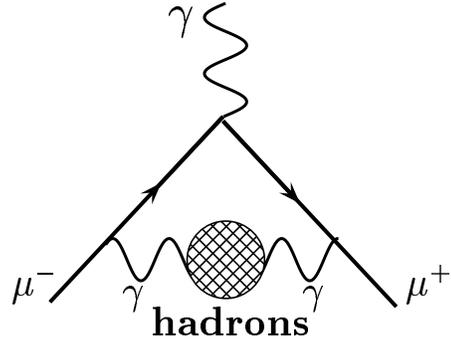}
\end{picture}

\vspace*{17mm}

\caption{Leading vacuum polarization contribution to $g-2$.}
\label{fig:fig3}
\end{figure}

The leading hadronic contribution to $g-2$ is given by the diagram
Fig.~\ref{fig:fig3}, where the blob represents the one-particle
irreducible contribution to the photon propagator.
To evaluate the diagram Fig.~\ref{fig:fig3} we can make use of the
representation (\ref{eq:DI}) for the vacuum polarization
function. Again one obtains a dispersion integral (Bouchiat, Michel
1961, Durand III 1962, Gourdin, de Rafael 1969)
\begin{eqnarray}
\begin{array}{c}{\bf
a_\mu \equiv \frac{g_\mu -2}{2} = \left(\frac{\alpha m_\mu}{3\pi}
\right)^2 \int\limits_{4 m_\pi^2}^{\infty}ds
\frac{R(s)\;\hat{K}(s)}{s^2}}
\end{array}
\label{eq:amuDI}
\end{eqnarray}
which may be evaluated in terms of the experimentally
accessible quantity
\begin{eqnarray*}
\begin{array}{c}{\bf
R(s)=\frac{\sigma_{tot}(e^+ e^- \rightarrow \gamma^* \rightarrow {\rm
hadrons})}{\sigma (e^+ e^- \rightarrow \gamma^* \rightarrow \mu^+
\mu^-)}}
\end{array}\;\;.
\end{eqnarray*}
\noi

\vspace*{-8mm}

\vspace*{-7.5mm}

\begin{figure}[h]
\begin{picture}(10,50)
\includegraphics{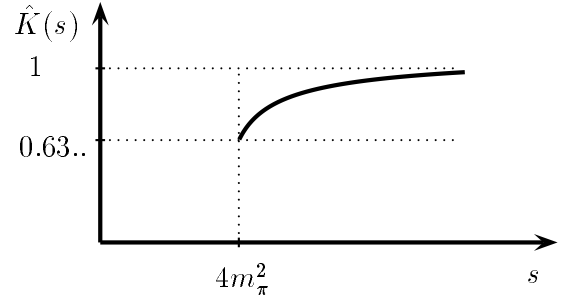}
\end{picture}

\vspace*{-8mm}

\caption{The kernel $\hat{K}(s)$ as a function of energy.}
\label{fig:fig4}
\end{figure}

The kernel $\hat{K}(s)$ of (\ref{eq:amuDI}), depicted in
Fig.~\ref{fig:fig4}, is a smooth {\sl bounded function}.

\vspace*{13mm}

%\newpage

\begin{table*}
\caption[]{
$a_{\mu}^{{\rm had}}\cdot 10^{10}$ and relative(rel) and
absolute(abs) errors in percent.}
\begin{tabular}{ccrcc}
\hline
 final state & range (GeV) & $a_{\mu}^{{\rm had}}\cdot 10^{10}$ &
 rel \% & abs \% \\
\hline
  $ \rho $   &(0.28, 0.81)    &426.66 ( 5.61) (10.62)& 2.8\% & 1.7\% \\
  $ \omega $ &(0.42, 0.81)    & 37.76 ( 0.45) ( 1.02)& 3.0\% & 0.2\% \\
  $ \phi   $ &(1.00, 1.04)    & 38.55 ( 0.54) ( 0.89)& 2.7\% & 0.1\% \\
  $ J/\psi $ &                &  8.60 ( 0.41) ( 0.40)& 6.7\% & 0.1\% \\
  $\Upsilon$ &                &  0.10 ( 0.00) ( 0.01)& 6.7\% & 0.0\% \\
hadrons      &(0.81, 1.40)    &112.85 ( 1.33) ( 5.49)& 5.0\% & 0.8\% \\
hadrons      &(1.40, 3.10)    & 56.43 ( 0.45) ( 7.22)&12.8\% & 1.0\% \\
hadrons      &(3.10, 3.60)    &  4.47 ( 0.23) ( 0.86)&19.9\% & 0.1\% \\
hadrons      &(3.60, 9.46)    & 14.06 ( 0.07) ( 0.90)& 6.5\% & 0.1\% \\
hadrons      &(9.46, 40.0)    &  2.70 ( 0.03) ( 0.13)& 4.9\% & 0.0\% \\
 QCD &(40.0, $\infty$ ) &  0.16 ( 0.00) ( 0.00)& 0.2\% & 0.0\% \\
\hline
   total     &                &702.35 ( 5.85) (14.09)& 2.2 \% & 2.2\% \\
\hline
\end{tabular}
\label{tab:amu}
\end{table*}

\begin{table}[h]
\vspace*{-3mm}
\caption[]{Various estimates of $a_\mu^{\rm had}\times 10^{10}$}

\begin{tabular}{ll}
\hline
 $a_{\mu}^{\rm had}\times 10^{10}$ & Author (Year) [Ref.] \\
\hline
650 ~~(50) & Gourdin, de Rafael (69) \ci{GdeR} \\
680 ~~(90) & Bramon, Etim, Greco (72) \ci{BEG72} \\
663 ~~(85) & Barger, Long, Olsen (75) \ci{Barger75} \\
702 ~~(80) & Calmet et al. (78) \ci{Narison78} \\
684 ~~(11) & Barkov et al. (85) \ci{Barkov85} \\
707~(6)(16) & Kinoshita et al. (85) \ci{Kino2} \\
710(10)(5)~ & Casas et al. (85) \ci{CLY} \\
705~(6)(5)~ & Martinovi\v{c}, Dubni\v{c}ka (90) \ci{MD} \\
724~(7)(26) & Jegerlehner (91) \ci{jegup} \\
699~(4)(2)~ & Dubni\v{c}kov\'{a} et al. (92) \ci{DD} \\
702~(6)(14) & Eidelman, Jegerlehner (95) \ci{EJ95}\\
711 ~~(10)  & Adel, Yndur\'ain (95) \ci{AY95} \\
702~(8)(13) & Worstell, Brown (95) \ci{WB95}\\
\hline
\end{tabular}
\label{tab:gm2stat}
\end{table}
\noi
Note that the energy denominator $1/s^2=1/E^4$ {\sl dramatically
enhances the non--perturbative low energy contribution}
\subsection{Results for $a_\mu^{\rm had}$:}

\vspace*{3mm}

\begin{figure}[h]
\begin{picture}(10,10)
\includegraphics{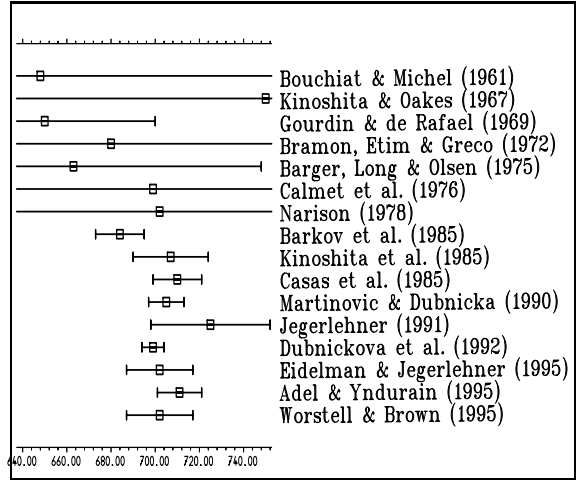}
\put(0,-174){\framebox(213,180)}
\end{picture}

\vspace*{51mm}

\caption{Evaluations of $a_{\mu}^{\rm hadrons}\times 10^{10}$ by
different authors.}
\label{fig:gm2stat}
\end{figure}

The present status od the hadronic vacuum contribution to $g-2$ is
summarized in Table~\ref{tab:gm2stat} and Fig.~\ref{fig:gm2stat}. Only
the leading hadronic vacuum polarization diagram is included. Our
estimate is  
\be
a_\mu^{\rm had}= 702.35 (15.28) \times 10^{-10}\;.
\ee
Our values for $a_{e}^{{\rm had}}$ and $a_{\tau}^{{\rm had}}$ are
given in Table~\ref{tab:aeatau}.\\[1cm]

\begin{table*}
\caption[]{Contributions to
$a_{e}^{{\rm had}}\cdot 10^{14}$ and $a_{\tau}^{{\rm had}}\cdot
10^{8}$.}
\begin{tabular}{ccrr}
\hline
 final state & range (GeV) &
$a_{e}^{{\rm had}}\cdot 10^{14}$ (stat) (syst) &
$a_{\tau}^{{\rm had}}\cdot 10^{8}$ (stat) (syst)\\
\hline
  $ \rho $   &(0.28, 0.81)    &118.57 ( 1.59) ( 2.97)&137.42 ( 1.61) ( 3.33)\\
  $ \omega $ &(0.42, 0.81)    & 10.07 ( 0.12) ( 0.27)& 14.62 ( 0.18) ( 0.39)\\
  $ \phi $   &(1.00, 1.04)    &  9.86 ( 0.14) ( 0.23)& 20.36 ( 0.29) ( 0.47)\\
  $ J/\psi $ &                &  2.04 ( 0.10) ( 0.09)& 12.80 ( 0.60) ( 0.61)\\
  $\Upsilon$ &                &  0.02 ( 0.00) ( 0.00)&  0.24 ( 0.01) ( 0.01)\\
hadrons      &(0.81, 1.40)    & 29.16 ( 0.35) ( 1.41)& 56.22 ( 0.62) ( 3.00)\\
hadrons      &(1.40, 3.10)    & 13.70 ( 0.11) ( 1.75)& 56.32 ( 0.54) ( 7.60)\\
hadrons      &(3.10, 3.60)    &  1.06 ( 0.06) ( 0.20)&  6.66 ( 0.35) ( 1.28)\\
hadrons      &(3.60, 9.46)    &  3.31 ( 0.02) ( 0.21)& 26.43 ( 0.13) ( 1.70)\\
hadrons      &(9.46, 40.0)    &  0.63 ( 0.01) ( 0.03)&  6.78 ( 0.06) ( 0.33)\\
 QCD &(40.0, $\infty$) &  0.04 ( 0.00) ( 0.00)&  0.45 ( 0.00) ( 0.00)\\
\hline
   total     &                &188.47 ( 1.65) ( 3.75)&338.30 ( 1.97) ( 9.12)\\
\hline
\end{tabular}
\label{tab:aeatau}
\end{table*}
\section{CONCLUSION AND OUTLOOK}
\noi
{{\liste\symbol{'340}}}~~Muon anomalous magnetic moment:
The present uncertainty due to hadronic vacuum polarization is
\be
(g-2)_\mu  &:& \delta a_\mu \sim 16 \times 10^{-10}
\ee
This can be reduced to $8 \times 10^{-10}$ by ongoing measurements at
VEPP 2M in Novosibirsk~\ci{Khazin94} and in a forthcoming experiment
at DA$\Phi$NE in Frascati~\ci{Franzini95}. This has to be compared
with the experimental precision of the forthcoming BNL experiment 
E821 at Brookhaven~\ci{BNL}. The present experimental precision is 
$84 \times 10^{-10}$ was obtained in the CERN experiment~\ci{amuexp}.
A look at Table~\ref{tab:amu} shows that a
measurement of $e^+e^- \ra$ hadrons for 1.5 GeV to 3.6 GeV to 1\%
accuracy is urgently needed in order to be able to fully exploit the
BNL measurement of $a_\mu$.\\ \noi {{\liste\symbol{'340}}}~~Bhabha
scattering: the hadronic uncertainty is not a limiting factor in small
angle Bhabha scattering (luminosity monitoring). The estimated
uncertainty is $\lsim 0.06$ \% .
\\ \noi
{{\liste\symbol{'340}}}~~Precision measurements at LEP and SCL:
the hadronic uncertainty is close to become the limiting factor for precision
measurements of some observables like 
\be
\sin^2 \Theta^\ell_{\rm eff} & : & \delta \sin^2\Theta \sim 0.00025
\ee
This compares to the present accuracy of $\delta \sin^2\Theta \sim
0.00030$ of LEP and SLC data.\\ \noi {{\liste\symbol{'340}}}~~In
principle one could try to do a direct measurement of $(g-2)_{\mu
\:\rm had}$ and $\az$. Unfortunately, in this case one would lose the
most important predictions and in addition the result would be model
dependent !.\\ \noi {{\liste\symbol{'340}}}~~Reliable theoretical
calculations are not in sight.\\ \noi {{\liste\symbol{'340}}}~~{A real
breakthrough in improving on the hadronic uncertainty will require
dedicated efforts in high precision measurements of $R(s)$ in a wide
energy range.}

\section{APPENDIX: Analytic expressions.}

In many cases, in some interval $4m_\pi^2 \leq s_1<s_2\leq \infty$,
one may approximate $R(s)$ by simple analytic expressions and the
dispersion integral
\be
\dal_{had}(s)&=&\frac{-s}{4 \pi^2 \alpha} {\cal P}\!\!\! \int_{s_1}^{s_2}\:ds'\:
\frac{\sigma_{had}(s')}{s'-s} \crn
             &=&\frac{\alpha}{3 \pi} {\cal P}\!\!\!
\int_{s_1}^{s_2}\:ds'\:
\left\{ \frac{1}{s'}-\frac{1}{s'-s} \right\} R(s') \;\;,
\ee
may be performed analytically.

We consider a few examples in the following.

\bit
\item {\bf Linear extrapolation}
\eit
For example, for non-resonant contributions (neighboring) data points
may be approximated by a linear function $$R(s)=a+b \sqrt{s}$$ in the
c.m. energy which gives a contribution\footnote{Eq. (7) of
Ref. \ci{Burkhardt89} contains several misprints}{\small
\be
\dal_{had}(s)&=&\frac{\alpha}{3 \pi}
\left\{ 2 a \ln \frac{W_2}{W_1}
- (a+b W) \ln \frac{W_2-W}{W_1-W}
 \right. \\ &&  \left.
- (a-b W) \ln \frac{W_2+W}{W_1+W} \right\} \crn
               &=& \frac{\alpha}{3 \pi}
\left\{ 2 a \ln \frac{W_2}{W_1} - a \ln |\frac{s_2-s}{s_1-s}|
- b f(W) \right\},
\ee }
where $W_i=\sqrt{s_i}$, $W=\sqrt{s}$ and {\small
\be
f(W)=
\left\{ \begin{array}{c}
  W\:(\ln |\frac{W_2-W}{W_1-W}|-\ln |\frac{W_2+W}{W_1+W}|);\; q^2=W^2>0 \\
 \!\! 2 |W|(\arctan\frac{|W|}{W_2}-\arctan\frac{|W|}{W_1});\;  q^2=W^2<0.
\end{array} \right.
\ee }

\bit
\item {\bf Narrow width resonance}
\eit
The contribution from a zero width resonance
\be
\sigma_{NW}(s)=\frac{12 \pi^2}{M_R} \Gamma_{e^+ e^-} \delta(s-M_R^2)
\ee
or
\be
R_{NW}(s)=\frac{9 \pi M_R}{\al^2} \Gamma_{e^+ e^-} \delta(s-M_R^2)
\ee
is given by
\be
\dal_{res}(s)=\frac{3 \Gamma_{e^+ e^-}}{\alpha M_R}\frac{s}{s-M_R^2}
\ee
which in the limit $|s|\gg M_R^2$ becomes
\be
\dal_{res}(s)=\frac{3 \Gamma_{e^+ e^-}}{\alpha M_R}\;\;.
\ee

\bit
\item {\bf Breit-Wigner resonance}
\eit
The contribution from a Breit-Wigner resonance
\be
\sigma_{BW}(s)=\frac{3 \pi}{s}
\frac{\Gamma \Gamma_{e^+ e^-}}{(\sqrt{s}-M_R)^2+\frac{\Gamma^2}{4}}
\ee
or
\be
R_{BW}(s)=\frac{9}{4 \al^2}
\frac{\Gamma \Gamma_{e^+ e^-}}{(\sqrt{s}-M_R)^2+\frac{\Gamma^2}{4}}
\ee
is given by
\be
\dal_{res}(s)=\frac{3 \Gamma \Gamma_{e^+ e^-}}{4 \pi \alpha}
\left\{ I(0)-I(W)
\right\}
\ee
where
\be \bary{l}
I(W) =\frac{1}{2ic} \\ 
\left\{\frac{1}{W-M_R-ic} 
\left( \ln \frac{W_2-W}{W_1-W}-\ln \frac{W_2-M_R-ic}{W_1-M_R-ic} \right)
\right. \\ \left. 
- \frac{1}{W+M_R+ic} 
\left( \ln \frac{W_2+W}{W_1+W}-\ln \frac{W_2-M_R-ic}{W_1-M_R-ic}
\right)
\right. \\ \left. 
- h.c. \right\} \eary 
\ee
with $c=\Gamma/2$. This may be written as
\be \bary{l}
\dal_{res}(s)=
\frac{3 \Gamma_{e^+ e^-}}{\pi \alpha M_R} \frac{M_R^2}{M_R^2+c^2} 
 \frac{1}{(s-M_R^2+c^2)^2+M_R^2 \Gamma^2} \\  
\left\{ s(s-M_R^2+3c^2) 
\right. \\ \left.                                     
\left(\pi-\arctan\frac{c}{W_2-M_R}-\arctan\frac{c}{M_R-W_1} \right)
 \right. \\
-\frac{\Gamma}{4M_R} \left[
   s(s-3M_R^2+c^2) \ln \frac{(W_2-M_R)^2+c^2}{(W_1-M_R)^2+c^2}
 \right. \\
+(s+M_R^2+c^2)(M_R^2+c^2) \ln |\frac{s_2-s}{s_1-s}|
         \\
+2 M_R \:(M_R^2+c^2) f(W)
         \\ \left. \left.
+2 \:((s-M_R^2+c^2)^2+M_R^2 \Gamma^2)\ln \frac{W_2}{W_1} \right] \right\}
\eary
\ee
with $f(W)$ given above.
For $W_1 \ll M_R \ll W_2$ and $\Gamma \ll M_R$ this
may be approximated by
\ba
\dal_{res}(s)=
\frac{3 \Gamma_{e^+ e^-}}{\alpha M_R}
 \frac{s(s-M_R^2+3c^2)}{(s-M_R^2+c^2)^2+M_R^2 \Gamma^2}
\ea
which agrees with Eqs. (3) and (4) in the limits $\Gamma^2 \ll |s-M_R^2|,M_R^2$
and $|s|\gg M_R^2$, respectively.

\bit
\item {\bf Breit-Wigner resonance: Field theory version}
\eit
Finally, we consider a field theoretic form of a Breit-Wigner resonance
obtained by the Dyson summation of a massive spin 1 transversal part of the
propagator in the approximation that the imaginary part of the self--energy
yields the width by Im$\Pi_V(M_V^2)=M_V \Gamma_V$ near resonance.

\be
\sigma_{BW}(s)=\frac{12 \pi}{M_R^2} \frac{\Gamma_{e^+ e^-}}{\Gamma}
\frac{s \Gamma^2 }{(s-M_R^2)^2+M_R^2 \Gamma^2}
\ee
which yields
\be \bary{l}
\dal_{res}(s)=\frac{3 \Gamma_{e^+ e^-}}{\pi \alpha M_R}
\frac{s(s-M_R^2-\Gamma^2)}{(s-M_R^2)^2+M_R^2 \Gamma^2} \\
\left\{ \left( \pi-\arctan\frac{\Gamma M_R}{s_2-M_R^2}
                  -\arctan\frac{\Gamma M_R}{M_R^2-s_1} \right) \right. \\
\left.
         -\frac{\Gamma}{M_R}\frac{s}{(s-M_R^2-\Gamma^2)}
        \left( \ln |\frac{s_2-s}{s_1-s}|
              -\ln |\frac{s_2-M_R^2-iM_R \Gamma}{s_1-M_R^2-iM_R \Gamma}| \right)
\right\} \eary
\ee
and reduces to
\be
\dal_{res}(s)=\frac{3 \Gamma_{e^+ e^-}}{\alpha M_R}
\frac{s(s-M_R^2-\Gamma^2)}{(s-M_R^2)^2+M_R^2 \Gamma^2}
\ee
for $s_1\ll M_R^2 \ll s_2$ and $\Gamma \ll M_R$. Again we have the known
limits for small $\Gamma$ and for large $|s|$. For broad resonances 
the different parametrizations of the resonance in general yield very
different results. Therefore, it is important to know how a resonance
was parametrized to get the resonance parameters like $M_R$ and $\Gamma$.
For narrow resonances as considered in our application results are
not affected in a relevant way by using different parametrizations.

%\newpage

\end{document}